\documentclass{mem}
\pdfoutput=1
\usepackage{txfonts}
\usepackage{graphicx}
\begin{document}
\def\teff{$T\rm_{eff }$}
\def\kms{$\mathrm {km s}^{-1}$}
\title{Photometric brown-dwarf classification}
\subtitle{A method to identify and accurately classify large
  samples of brown dwarfs without spectroscopy}
\author{
N. \,Skrzypek\inst{1} 
\and S.J. \, Warren\inst{1}}
  \offprints{N. Skrzypek}
\institute{Astro Group, Imperial College London, Blackett Laboratory, Prince Consort Road, SW7 2AZ, London
\email{n.skrzypek10@imperial.ac.uk}}
\titlerunning{Photo-type brown dwarf classification}
\abstract{We have developed a method {\em ``photo-type''} to identify
  and accurately classify L 
  and T dwarfs, onto the standard system, from photometry alone. We combine SDSS, UKIDSS and
  WISE data and classify point sources by comparing the izYJHKW1W2
  colours against template colours for quasars, stars, and brown
  dwarfs. In a sample of $6.5\times10^6$ bright point
  sources, J$<$17.5, from 3150 deg$^2$, we identify and type 898 L and
  T dwarfs, making this the largest homogeneously selected sample of
  brown dwarfs to date. The sample includes 713 (125) new (previously
  known) L dwarfs and 21 (39) T dwarfs. For the previously-known
  sources, the scatter in the plot of photo-type vs spectral type
  indicates that our photo-types are accurate to 1.5 (1.0) sub-types rms for
  L (T) dwarfs. Peculiar objects and candidate unresolved
  binaries are identified.  } \maketitle{}
\section{Introduction}
Brown dwarfs were discovered less than twenty years ago, but the number known is growing rapidly, and the total number of spectroscopically classified LTY dwarfs catalogued at DwarfArchives.org recently surpassed 1000. This sample is heterogeneous, culled from several surveys with different characteristics, particularly SDSS, 2MASS, UKIDSS, and WISE, using search strategies that at each stage have explored new parameter space. The temperature sequence of sub-stellar objects is now mapped all the way down to $\sim~400$K.  Individual large samples include those of Kirkpatrick et al. (2000, 2011), Hawley et al. (2002), Burningham et al. (2010, 2013), Schmidt et al. (2010), and Mart{\'{\i}}n et al. (2010). All these samples explore or have been optimised for a limited section of the brown dwarf spectral sequence, e.g. primarily L dwarfs from SDSS, T dwarfs from UKIDSS, and late T dwarfs and Y dwarfs from WISE. Some of these samples are themselves heterogeneous, and so not suitable for statistical analysis \---\ which of course is understandable in the phase of exploring a new population. Spectroscopic classification requires substantial resources. For example, at J$\sim 17.5$, something like 30min on an 8m class telescope is required for a good spectrum. \\
A good example of the state of the art in deriving a statistical sample of brown dwarfs is the study of the sub-stellar birth rate by Day-Jones et al. (2013). They selected candidate mid-L to mid-T dwarfs, using well-defined selection criteria, and discovered 63 new brown dwarfs brighter than J=18.1, using XShooter spectroscopy on the VLT. To make a substantial step forward in this type of work we need to produce statistical samples that are an order of magnitude larger. A large sample is needed to reduce the statistical errors, but may also be used to quantify the variation in properties for any spectral subclass, and to discover rare types, by identifying outliers.\\ 
This paper describes an alternative search and classification method that starts from existing survey data, SDSS+UKIDSS+WISE, and exploits the wide wavelength range to determine accurate spectral types without the need for spectroscopy.
\section{Method}
The basis of the method is to compare the multiwavelength izYJHKW1W2 photometry
from SDSS+UKIDSS+WISE against the colours of previously classified L
and T dwarfs, in the form of polynomial relations of colour against
spectral type\footnote{See Aberasturi et al. (2011) for an earlier search for brown dwarfs using SDSS+2MASS+WISE.}. In this way the classification is ultimately tied to
the templates that define spectral types for L and T dwarfs. UKIDSS lies wholly within SDSS, and WISE is all-sky, so the common
overlap of these three surveys is currently the best multi-wavelength
dataset for a search for brown dwarfs. We computed that, brighter than J=17.5, nearly all 
spectral types L0 to T8 will be detected
in all the bands Y, J, H, K (UKIDSS) and W1, W2 (WISE), as well as in
at least one of the SDSS $i$ or $z$ bands (meaning that a photometric
measurement will exist in both bands). The exceptions, discussed
later, are the latest T dwarfs near the sample limit. The brown-dwarf
search is one element of a larger programme to classify all bright
point sources in UKIDSS into categories star, brown dwarf and
quasar. \\
We found it simpler to execute the search in two stages. Starting with
UKIDSS DR9 we selected all point sources detected in YJHK, with a
match in SDSS within 10\arcsec (the fraction of sources lost due to high proper motion is tiny). This initial catalogue contained
6.5$\times10^6$ sources over 3150 deg$^2$. We classified all sources
by min-$\chi^2$ fitting against the template library. Then for all
sources that were classified as cool stars of class M6 or later we matched to WISE in order
to extract the W1 and W2 photometry. We then reclassified all
sources using the complete izYJHKW1W2 data. The quasar and star
templates are taken from Hewett et al. (2006), but we found that the
brown-dwarf templates are deficient in some respects, and so we
produced our own. We identified 130 spectroscopically classified L and T dwarfs in
DwarfArchives in our catalogue, supplemented with late M stars from
SDSS, and then fit polynomials to the variation of the seven colours
i-z, z-Y, Y-J, J-H, H-K, K-W1, W1-W2 against spectral type. The value
of the fit to each colour at each spectral type then defined the
templates. For each source we record the spectral type that provides
the min-$\chi^2$ fit over all the templates, and the value of
$\chi^2$.  The final sample contains 838 L dwarfs (of which 125 are
previously known) and 60 T dwarfs (39 previously known).
\begin{figure}
\resizebox{6.0cm}{!}{\includegraphics[clip=true]{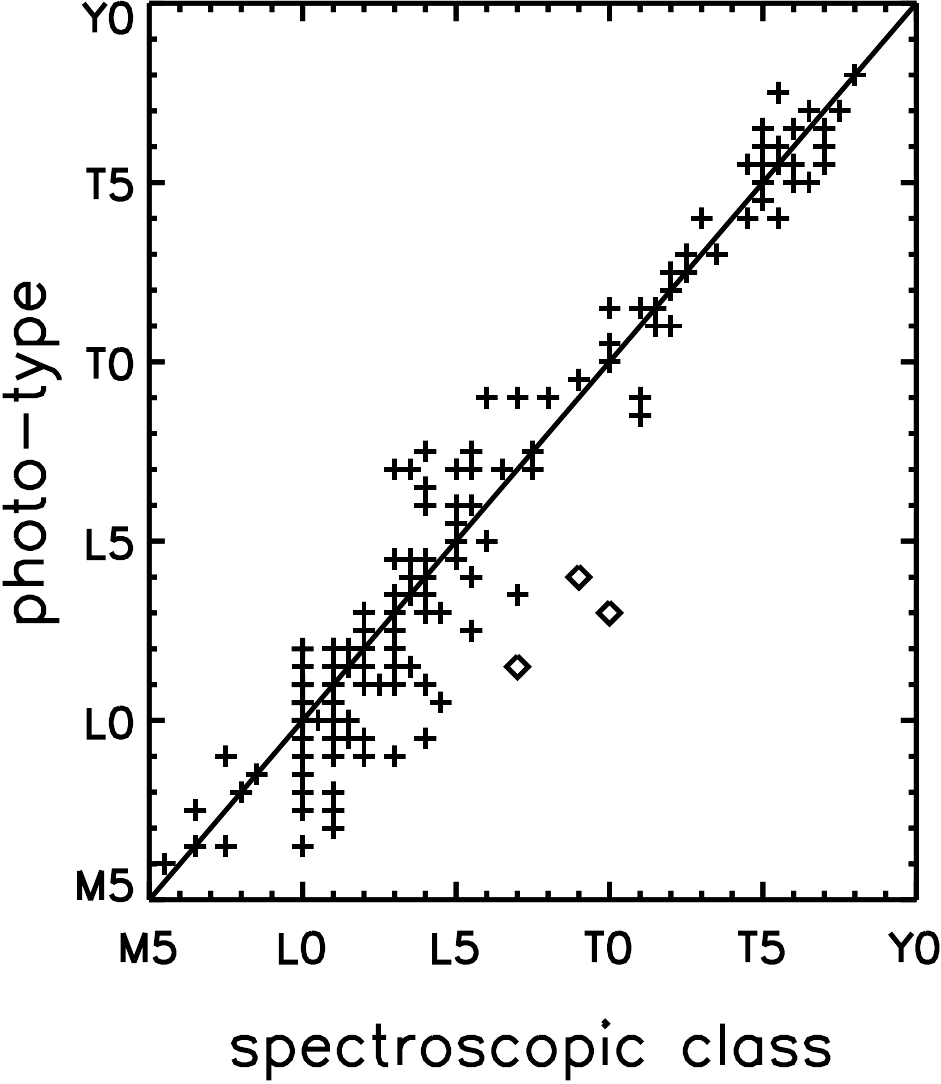}}
\caption{Plot of {\em ``photo-type''} class against spectroscopic class. This
  scatter indicates an accuracy of 1.5 spectral types rms for the L
  dwarfs and 1.0 spectral types rms for the
  T dwarfs.\footnotesize}
\label{accuracy}
\end{figure}
We can gauge the accuracy of the spectral type using the known
objects. In Fig. \ref{accuracy} we plot the {\em ``photo-type''}
classification against the spectroscopic
classification in DwarfArchive. Our L- and T-dwarf
classifications are accurate to 1.5 and 1.0 rms sub-classes
respectively. The 3 outliers (diamonds) are potential binaries
currently under investigation. It is possible that the classification
may be even more accurate, because part of the scatter in the plot
comes from the spectroscopic classification, due to the limited
wavelength coverage of most spectra. So far we have obtained four
follow-up spectra and all classifications agree within one sub-class.
\section{Completeness}
\begin{figure}
\resizebox{6.0cm}{!}{\includegraphics[clip=true]{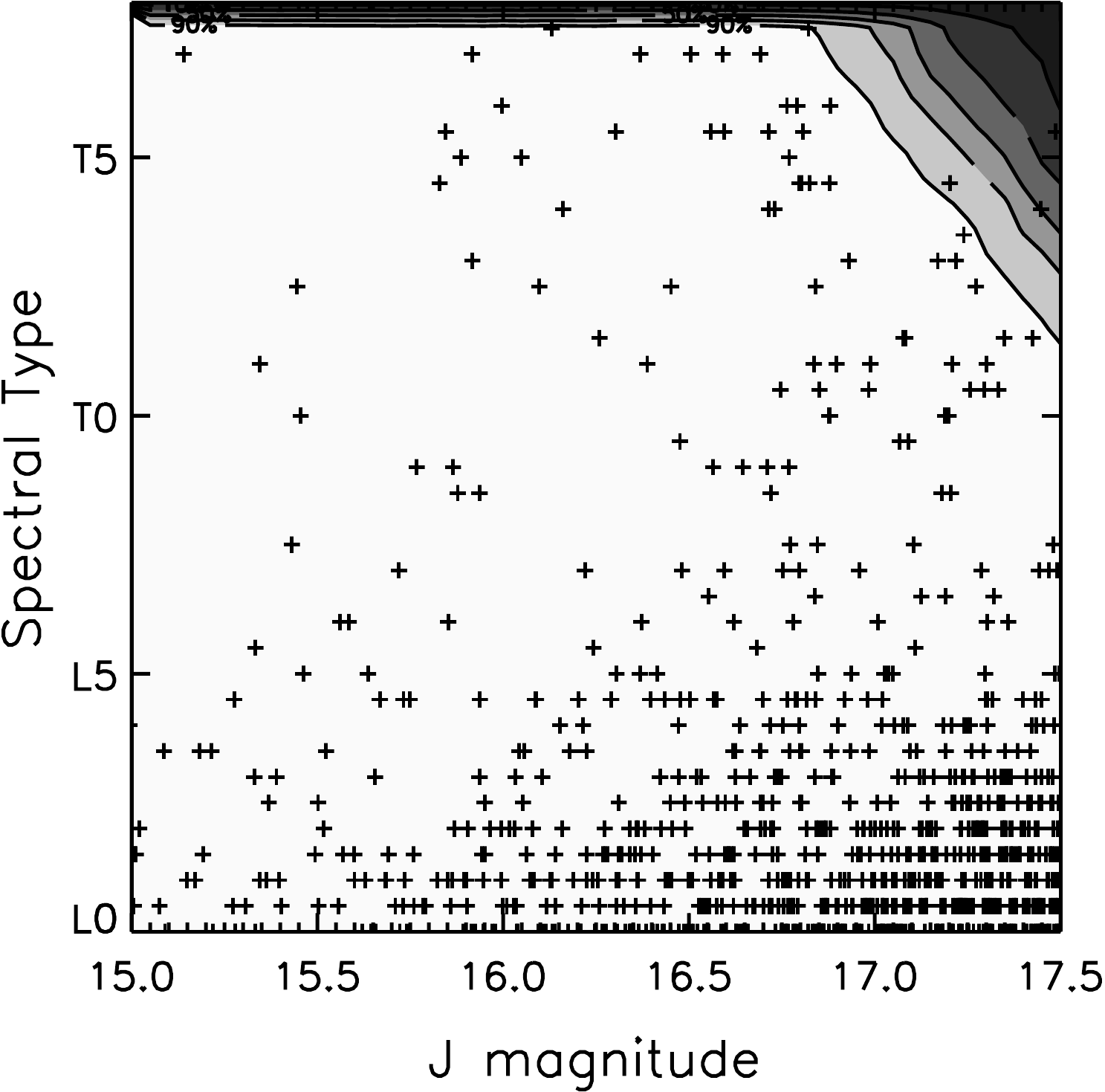}}
\caption{Plot illustrating the completeness of our sample, with the
  838 L and 60 T dwarfs overplotted as black crosses.\footnotesize}
\label{completeness}
\end{figure}
To quantify the completeness of the sample we performed Monte Carlo
simulations, creating synthetic colours for each spectral type, over
the magnitude range of the search, adding realistic errors, and
accounting for the detection limits in each band. The results are
shown in Fig. \ref{completeness}, showing that we only miss the
coolest T dwarfs, close to the sample magnitude limit. This is
primarily because as we approach J=17.5 the coolest T dwarfs fall
below the SDSS detection limit, and in some cases the UKIDSS K-band
limit. It would be possible to include these sources in principle, by
performing aperture photometry on the original images. However
the effort was not felt justified at this stage bearing in mind that
the UKIDSS dataset has already been searched extensively for late T dwarfs by several teams, e.g. Burningham et al. (2013). 
\section{Discussion}
We have presented a new method to identify and classify L and T dwarfs
from SDSS+UKIDSS+WISE data, which provides quite accurate
classifications without the need for follow-up spectroscopy. We
consider ``photo-type'' to be analogous to photometric redshifts for
galaxies (``photo-z''), and that it could be an easier route to
classification for some statistical applications. The $\chi^2$ measure
for each object quantifies the goodness of fit. We are investigating
sources with large $\chi^2$ in two ways. In one case we test whether a
binary solution provides a significantly better fit. In this way we
may be able to quantify the fraction of binaries of different
spectral-type combinations. Sources where a binary solution does not
improve the fit significantly, and the $\chi^2$ remains high, may be
peculiar and follow-up spectroscopy is planned for these sources. 
\begin{acknowledgements}
We are grateful to Jackie Faherty for obtaining four spectra for this programme, and to Subhanjoy Mohanty for helpful comments.
\end{acknowledgements}
{}
\end{document}